\documentclass[10pt,final]{iopart}

\usepackage{times}
\usepackage{graphicx}
\usepackage{amssymb}
\usepackage{iopams}
\usepackage{cite}

\begin{document}

\title[Self-similar analytical model of the plasma expansion in a magnetic
field]{Self-similar analytical model of the plasma expansion in a magnetic field}

\author{H B Nersisyan$^{1,2}$, K A Sargsyan$^1$, D A Osipyan$^1$, M V Sargsyan$^1$
and H H Matevosyan$^1$}

\address{$^1$ Division of Theoretical Physics, Institute of Radiophysics and Electronics,
0203 Ashtarak, Armenia}
\address{$^2$ Centre of Strong Fields Physics, Yerevan State University, Alex Manoogian
str. 1, 0025 Yerevan, Armenia}

\ead{hrachya@irphe.am}

\begin{abstract}
The study of hot plasma expansion in a magnetic field is of interest for many astrophysical applications.
In order to observe this process in laboratory, an experiment is proposed in which an ultrashort laser
pulse produces a high-temperature plasma by irradiation of a small target. In this paper an analytical
model is proposed for an expanding plasma cloud in an external dipole or homogeneous magnetic field. The
model is based on the self-similar solution of a similar problem which deals with sudden expansion of
spherical plasma into a vacuum without ambient magnetic field. The expansion characteristics of the plasma
and deceleration caused by the magnetic field are examined analytically. The results obtained can be used
in treating experimental and simulation data, and many phenomena of astrophysical and laboratory significance.
\end{abstract}

\submitto{\PS}

\section{Introduction}
\label{sec:1}

The problem of sudden expansion of hot plasma into a vacuum in the presence of an external magnetic field has
been intensively studied in the mid-1960s in connection with the high-altitude nuclear explosions. It has also
been discussed in the analysis of many astrophysical and laboratory applications (see, e.g., \cite{zak03,win05}
and references therein). Such kind of processes arise during the dynamics of solar flares and flow of the solar
wind around the Earth's magnetosphere, in active experiments with plasma clouds in space, and in the course of
interpreting a number of astrophysical observations \cite{zak03,win05,pon89,osi03,ner09}. Researches on this
problem are of considerable interest in connection with the experiments on controlled thermonuclear fusion
\cite{sgr76} (a review \cite{zak03} summarizes research in this area over the past four decades).

The expanding plasma pushes magnetic field out which is a consequence of magnetic flux conservation. Due to the
plasma expansion the surface of the given magnetic flux tube inside plasma increases. The conservation of the
magnetic flux through the surface of the tube leads to a decrease of the local magnetic field and the formation
of the diamagnetic cavity. Thus even if a magnetic field is nonzero initially inside the plasma it tends to zero
during plasma expansion. In addition, the plasma is shielded from the penetration of the external
field by means of surface currents circulating inside the thin layer on the plasma boundary. Ponderomotive forces
resulting from interaction of these currents with the magnetic field would act on the plasma surface as if there
were magnetic pressure applied from outside. Thus after some period of accelerated motion, plasma gets decelerated
as a result of this external force acting inward. The plasma has been considered as a highly conducting media with
zero magnetic field inside. From the point of view of electrodynamics it is similar to the expansion of a superconductor
in a magnetic field. An exact self-similar analytic solution for a uniformly expanding, highly conducting plasma
sphere in an external uniform and constant magnetic field has been obtained in \cite{kat61}. The non-relativistic
limit of this theory has been used by Raizer \cite{rai63} to analyze the energy balance (energy emission and
transformation) during the plasma expansion. The similar problem has been considered in \cite{dit00} within one-dimensional
geometry for a plasma layer. In our recent papers \cite{ner06} and \cite{ner10} we obtained an exact self-similar
analytic solution for the uniform relativistic expansion of the highly conducting plasma sphere or cylinder in the
presence of a dipole or homogeneous magnetic field, respectively.

As mentioned above the previous treatments \cite{dit00,kat61,rai63,ner06,ner10} were obtained assuming somewhat
idealized situation: uniform expansion, zero magnetic field and thermal pressure inside plasma, etc. Obviously such
simplified models are not capable to describe correctly the plasma expansion dynamics where an essential deceleration
(or acceleration) of the plasma boundary may occur. More realistic models for the plasma self-similar expansion have
been developed, for instance, in \cite{and80,dal11,hir11} (see also references therein) for planar (one-dimensional)
\cite{and80} and cylindrical \cite{dal11,hir11} expansions employing ideal magnetohydrodynamic (MHD) equations. It
should be noted that the ideal MHD is well justified since the typical experimental parameters are such that the expanding
plasma is collisionless and the dissipative effects are negligible \cite{zak03,win05,pon89,osi03,ner09}.

In the present paper we study the expansion of the spherical plasma cloud in the presence of a dipole or homogeneous
magnetic field taking into account the thermal effects. We employ the self-similar solution of a similar problem which
deals with expansion of spherical plasma into a vacuum without magnetic field. Unlike the previous studies \cite{and80,dal11,hir11}
we concentrate here on the dynamics (deceleration and/or acceleration) of the sharp plasma boundary. We found an analytical
solution which can be used in analyzing the recent experimental and simulation data (see, e.g., \cite{zak03,win05} and
references therein).

\section{Theoretical model}
\label{sec:2}

Usually the motion of the expanding plasma boundary is approximated as the motion with constant velocity
(uniform expansion). In the present study a quantitative analysis of plasma dynamics is developed on the
basis of one-dimensional spherical radial model. Within the scope of this analysis the initial stage of
plasma acceleration, later stage of deceleration and the process of stopping at the point of maximum
expansion are examined.

We consider a collisionless magnetized plasma expanding into vacuum. The relevant equations governing the
expansion are those of ideal MHD \cite{lan84} assuming that the characteristic length scales for the plasma
flow are much larger than the Debye length and the Larmor radius of the ions. Thus
\begin{eqnarray}
\frac{\partial \rho }{\partial t}+\nabla \cdot \left( \rho \mathbf{v}\right) =0 , \nonumber \\
\rho \left[ \frac{\partial \mathbf{v}}{\partial t}+\left( \mathbf{v}\cdot
\nabla \right) \mathbf{v}\right]  =\frac{1}{4\pi }\left[ \nabla \times
\mathbf{H}\right] \times \mathbf{H}-\nabla P , \nonumber \\
\frac{\partial \mathbf{H}}{\partial t} =\nabla \times \left[ \mathbf{v} \times \mathbf{H}\right] , \quad
\nabla \cdot \mathbf{H} =0 , \label{eq:a1}
\end{eqnarray}
where $\rho$, $\mathbf{v}$, $\mathbf{H}$, and $P$ are the mass density, the velocity, the magnetic field, and
the pressure respectively. These equations must be accompanied by the adiabatic equation of state $P \sim \rho^%
{\gamma}$ ($\gamma =C_{p}/C_{V}>1$ is the adiabatic index, $C_{p}$ and $C_{V}$ are the heat capacities at constant
pressure and constant volume, respectively) and the equation of conservation of entropy, which expresses the fact
that the plasma dynamics is adiabatic in the absence of dissipation. Using the thermodynamic relation between
entropy, pressure, and internal energy as well as equation (\ref{eq:a1}) the equation for pressure reads \cite{zel02}
\begin{equation}
\frac{\partial P}{\partial t}+(\mathbf{v}\cdot \nabla ) P +\gamma P \left(\nabla \cdot \mathbf{v}\right) =0 .
\label{eq:a2}
\end{equation}

The equation of conservation of energy is derived from the system of equations (\ref{eq:a1}) and is given by
\cite{lan84}
\begin{equation}
\frac{\partial \varepsilon }{\partial t}+\nabla \cdot \mathbf{Q}=0 ,
\label{eq:a3}
\end{equation}
where $\varepsilon$ and $\mathbf{Q}$ are the energy and energy flux densities, respectively, with
\begin{eqnarray}
\varepsilon &=&\frac{\rho v^{2}}{2}+\frac{P}{\gamma -1}+\frac{H^{2}}{8\pi } ,  \nonumber \\
\mathbf{Q} &=&\mathbf{v}\left(\frac{\rho v^{2}}{2} +\frac{\gamma P}{\gamma -1}\right) +\frac{1}{4\pi }\left[
\mathbf{H}\times \left[ \mathbf{v}\times \mathbf{H}\right] \right] .  \label{eq:a4}
\end{eqnarray}
Here the energy $\varepsilon$ consists of the kinetic (first term), the internal (second term) and the magnetic
field (third term) energies. The last term in the energy flux density $\mathbf{Q}$ represents the Poynting vector
(let us recall that in ideal MHD the electric field is given by $\mathbf{E} =-\frac{1}{c}[\mathbf{v}\times \mathbf{H}]$).

With the theoretical basis presented so far, we now take up the main topic of this paper. This is to study the
dynamics of the plasma boundary expanding in an ambient magnetic field.
Consider the magnetic dipole $\mathbf{p}$ and a plasma spherical cloud with radius $a(t)$ located at the
origin of the coordinate system. The dipole is placed in the position $\mathbf{r}_{0}$ from the center
of the plasma cloud ($a(t) < r_{0}$). The orientation of the dipole is given by the angle $\theta_{p}$
between the vectors $\mathbf{p}$ and $\mathbf{r}_{0}$. We denote the strength of the magnetic field of
the dipole by $\mathbf{H}_{0}(\mathbf{r})$. The energy, which is transferred from plasma to electromagnetic
field is the mechanical work performed by the plasma on the external magnetic pressure $H_{0}^{2}(\mathbf{r})/8\pi$.
Taking into account this effect and the energy conservation (\ref{eq:a3}) integrated over the spherical plasma
volume $\Omega_{p}$ (with $0 \leqslant r\leqslant a(t)$) the equation of balance of plasma energy is as follows:
\begin{eqnarray}
\frac{4\pi}{\gamma -1} \int_{0}^{a\left( t\right)}P(r,t) r^{2}dr +2\pi \int_{0}^{a\left( t\right)}
\rho (r,t) v^{2}(r,t) r^{2}dr  \label{eq:1}  \\
+\int_{\Omega}\frac{H_{0}^{2}\left( \mathbf{r}\right) }{8\pi } d\mathbf{r}=W_{0} ,   \nonumber
\end{eqnarray}
where $\Omega $ is the volume of the spherical shell $a_{0}\leqslant r\leqslant a(t)$, $a_{0}= a(0)$
($a(t)\geqslant a_{0}$) and $W_{0}$ are the initial radius and energy of the plasma. When the plasma cloud is
introduced into a background magnetic field, the plasma expands and excludes the background magnetic field to
form a magnetic cavity. The magnetic energy of the dipole in the excluded volume
is represented by the last term in equation~(\ref{eq:1}). Initial plasma velocity is supposed to be $v(r,0) =v_{m}%
(r/a_{0})$ at $r\leqslant a_{0}$ and $v(r,0) =0$ at $r> a_{0}$, where $v_{m}$ is the initial velocity of the
plasma boundary ($v_{m} =\dot{a}(0)$).

The obtained energy balance equation can be effectively used if profiles of velocity $v(r,t)$, pressure $P(r,t)$,
and mass density $\rho (r,t)$ are known functions of the plasma radius $a(t)$. We will take these dependencies from
the solution of a similar problem which deals with sudden expansion of spherical plasma into a vacuum without
ambient magnetic field \cite{zel02}. The simplest class of solutions available in this case are so-called self-similar
solutions. They are realized under the specified initial conditions. We will set the initial conditions with a parabolic
distribution of pressure and mass density, which describe hot and dense initial plasma state with sharp boundary
localized at $r=a_{0}$. The self-similar solutions are characterized by a velocity distribution linearly dependent
on $r$ (see, e.g., \cite{zel02}). At $r\leqslant a(t)$
\begin{equation}
v\left( r,t\right) =r\frac{\dot{a}\left( t\right) }{a\left( t\right) } ,
\label{eq:2}
\end{equation}
where unknown $a(t)$ is the radius of sharp plasma boundary while $\dot{a} (t)$ is the velocity of the boundary.
The specification of the mass density profile at $r\leqslant a(t)$ is given by
\begin{equation}
\rho \left( r,t\right) =\frac{\Gamma \left(\frac{5}{2}+q\right)}{\pi^{3/2}
\Gamma\left(1+q\right)} \frac{M}{a^{3}\left( t\right) } \left[1 -\frac{r^{2}}{a^{2}(t)}\right]^{q}
\label{eq:3}
\end{equation}
and equation~(\ref{eq:2}) for the velocity, automatically satisfies the continuity equation in (\ref{eq:a1}) for an
arbitrary function $a(t)$ and for an arbitrary parameter $q$. Here $M= \mathrm{const}$ is the total mass of plasma
cloud and $\Gamma (z)$ is the Euler function. Substitution of $v(r,t)$ into the entropy equation (\ref{eq:a2}) gives
at $r\leqslant a(t)$ the following solution for the pressure
\begin{equation}
P\left( r,t\right) =p_{\max }\left[ \frac{a_{0}}{a\left( t\right) }\right]
^{3\gamma }\left[ 1-\frac{r^{2}}{a^{2}\left( t\right) }\right]^{s} ,
\label{eq:4}
\end{equation}
where $s$ is an arbitrary parameter and $p_{\max}$ is the thermal pressure at the center of the spherical plasma cloud
at $t=0$. In addition the quantities $v(r,t)$, $\rho (r,t)$ and $P(r,t)$ vanish at $r>a(t)$, $v(r,t) =\rho (r,t) =P(r,t)=0$.
Substituting (\ref{eq:3}) and (\ref{eq:4}) into the fluid equation of motion (\ref{eq:a1}) and neglecting the term involving
the magnetic field $\mathbf{H}$ yields a second-order differential equation governing the motion of the plasma boundary $a(t)$.
The problem considered is not isentropic in general except the case when $s= q\gamma$. In the latter case of the isentropic
expansion the equation of state is given by $P \rho ^{-\gamma}= \mathrm{const}$. Throughout in this paper we will assume that
$q\geqslant 0$ and $s\geqslant 0$. The self-similar solution given by equations (\ref{eq:2})-(\ref{eq:4}) has been considered
by many authors. We refer to \cite{zel02} (see also the references therein) for details and for some applications of this
solution.

There are several limitations and shortcomings of the self-similar model considered above.
It should be emphasized that in general the plasma expansion process with and without ambient magnetic field is not self-similar
since there is a characteristic length scale, that is, the initial radius of the plasma $a_{0}$. However, during the later stage
(when $a(t)\gg a_{0}$) the plasma expansion asymptotically approaches to the self-similar regime since the role of the radius
$a_{0}$ becomes less important. This property is supported by the asymptotic solution of the system of partial differential
equations (\ref{eq:a1}) and (\ref{eq:a2}) \cite{zel02,sta60} as well as by the numerical simulations \cite{zel02} for the free
expansion (i.e. at $\mathbf{H} =0$). In addition, it has been shown that at the beginning stage of the expansion the deviation
of the solution (\ref{eq:2})-(\ref{eq:4}) from non-similar one is small and is additionally reduced due to the spatial
average in the equation of energy balance (\ref{eq:1}).

Equations~(\ref{eq:2})-(\ref{eq:4}) are an exact solution in the case of expansion into a vacuum without magnetic field.
However, equation~(\ref{eq:4}) does not satisfy the boundary condition, $P(a(t), t) =H^{2}_{0}/8\pi$, which is imposed in
the case of expansion into an ambient magnetic field. Therefore, in the present case with nonzero magnetic field for the
validity of the equations (\ref{eq:2})-(\ref{eq:4}) most critical is the domain close to the boundary of the plasma,
$r\simeq a(t)$. On the other hand if the magnetic pressure is smaller than the plasma average pressure $\bar{P}$,
$P_{\mathrm{mag}}\ll \bar{P}$, the difference between the exact solution in the magnetic field and free expansion model
is small and is localized in a narrow area near the surface of the cloud. These deviations are additionally reduced
due to the integration in the equation of energy balance. Estimating accuracy of the free expansion model, one should take
into account that the long stage of plasma deceleration corresponds to a high expansion ratio, $a(t)/a_{0}\gg 1$. Average
plasma pressure drops significantly and plays no role in energy balance equation (\ref{eq:1}) during this stage. In
accordance with the above boundary condition local pressure near the plasma edge must be equal to the magnetic
pressure outside. It causes deviation from the profile equation (\ref{eq:3}) and accumulation of plasma in this
area. This is confirmed independently by the numerical simulations \cite{gol78}. In the limiting case when all plasma
is localized near the front, one can expect an increase of the kinetic energy and longer stage of plasma deceleration
as compared with the self-similar expansion model.

Another critical domain for the violation of the self-similar solution (\ref{eq:2})-(\ref{eq:4}) is the final stage of
expansion when the plasma is fully stopped by the magnetic field. As mentioned above the average plasma pressure is
strongly reduced compared to the magnetic pressure and the equations (\ref{eq:2})-(\ref{eq:4}) clearly become invalid in
this case. However, if the critical time interval $\Delta t$, where $\bar{P} <P_{\mathrm{mag}}$, is much smaller than the
typical time scale of the plasma flow (up to the full stop) the contribution of this interval to the overall plasma
dynamics is negligible and the use of the self-similar solution is justified.

In the case of dipole magnetic field the volume integral in the last term of equation~(\ref{eq:1}) has been evaluated in
\cite{ner06}. The result reads
\begin{equation}
\int_{\Omega} \frac{H_{0}^{2}\left(
\mathbf{r}\right) }{8\pi }d\mathbf{r}=\frac{p^{2}}{32r_{0}^{3}}\left[
Q\left( \eta x\left( t\right) \right) -Q\left( \eta \right) \right] ,
\label{eq:5}
\end{equation}
where $\eta =a_{0}/r_{0}<1$, $x(t) =a(t)/a_{0}$ (note that $a(t)/r_{0}=\eta x(t) <1$), and
\begin{eqnarray}
&&Q\left( \eta \right) =\frac{1}{\left( 1-\eta ^{2}\right) ^{3}}
\left[\eta \left( 1-\eta ^{4}\right) \left( 3\cos
^{2}\theta _{p}-1\right) \right.   \label{eq:6}  \\
&&\left. +8\eta ^{3}\left( 1+\cos ^{2}\theta _{p}\right) \right]
-\frac{3\cos ^{2}\theta _{p}-1}{2}\ln \frac{1+\eta }{1-\eta } .   \nonumber
\end{eqnarray}%

Substituting equations~(\ref{eq:2})-(\ref{eq:4}) into (\ref{eq:1}) and integrating over $r$ yields first-order differential
equation for $a(t)$, which already satisfies initial condition $\dot{a}(0) =v_{m}$,
\begin{equation}
\dot{x}^{2}(\tau ) +\frac{\beta }{x^{3\left( \gamma -1\right)}}+\alpha \left[ Q\left( \eta x\left( \tau \right)
\right) -Q\left( \eta \right) \right] =1 .
\label{eq:7}
\end{equation}
Here two dimensionless quantities are introduced
\begin{equation}
\alpha =\frac{p^{2}}{32W_{0}r_{0}^{3}},  \quad   \beta =\frac{\pi^{3/2}
p_{\max} a_{0}^{3}}{\left( \gamma -1\right) W_{0}} \frac{\Gamma (1+s)}{\Gamma (\frac{5}{2}+s)} <1
\label{eq:8}
\end{equation}
which determine the magnetic and the thermal energies, respectively, in terms of the total initial energy $W_{0}$.
The latter is easily obtained from equation~(\ref{eq:1}) and reads
\begin{equation}
W_{0}=\frac{3Mv_{m}^{2}}{2(5+2q)} +\frac{\pi^{3/2} p_{\max} a_{0}^{3}}{\gamma -1 }
\frac{\Gamma (1+s)}{\Gamma (\frac{5}{2}+s)}.
\label{eq:9}
\end{equation}
From equations~(\ref{eq:8}) and (\ref{eq:9}) it is seen that $\beta <1$. New dimensionless variables are introduced as
follows: $x(\tau )=a(t)/a_{0}$, $\tau =t/t_{0}$, $t_{0}=a_{0}/u_{m}$, where $u_{m} =[2(5+2q) W_{0}/3M]^{1/2}$ is the
velocity of plasma expansion, achieved asymptotically at $t\to \infty$ in the case of expansion into a vacuum without
magnetic field (i.e. at $\alpha =0$).

The total energy of the plasma cloud at time $t$ is obtained from equation~(\ref{eq:7})
\begin{equation}
W\left( t\right) =W_{0}-\frac{p^{2}}{32r_{0}^{3}}\left[ Q\left(
\eta x\left( t\right) \right) -Q\left( \eta \right) \right] .
\label{eq:10}
\end{equation}
Note that the function $Q(\eta )$ monotonically increases with the argument and the plasma cloud energy decreases
with time.

Consider also the case of uniform magnetic field when $\mathbf{H}_{0}=\mathrm{const}$. In this case the
volume integral in equation~(\ref{eq:5}) is replaced by $(H^{2}_{0}/6) (a^{3}(t)-a^{3}_{0})$ and the differential
equation (\ref{eq:7}) for the plasma boundary reads
\begin{equation}
\dot{x}^{2}(\tau ) + \frac{\beta }{x^{3\left( \gamma -1\right) }}+\sigma [x^{3}(\tau ) -1]=1 ,
\label{eq:11}
\end{equation}%
where $\sigma =W_{\mathrm{mag}}/W_{0}$, $W_{\mathrm{mag}}=(4\pi a_{0}^{3}/3)P_{\mathrm{mag}}$ is the initial
magnetic energy in the plasma volume, and $P_{\mathrm{mag}}=H_{0}^{2}/8\pi $ is the magnetic field pressure.

Equations (\ref{eq:7}) and (\ref{eq:11}) coincide with the equation of the one-dimensional motion of the
point-like particle in the potential $U(x)$ which is determined by second and third terms of equations~(\ref{eq:7})
and (\ref{eq:11}). The distance $x_{s}$ of the plasma cloud motion up to the full stop (the stopping length) at
the turning point is determined by $U(x_{s}) =1$. In particular, it is easier to obtain the stopping length
in the case of homogeneous magnetic field and at vanishing thermal pressure ($\beta =0$). Then from equation~(\ref{eq:11})
one obtains the equation of motion
\begin{equation}
\dot{x}^{2} = 1+\sigma -\sigma x^{3} .
\label{eq:12}
\end{equation}
It is seen that in this case the stopping length is given by $x_{s} =(1+1/\sigma )^{1/3}$. The solution of
equation~(\ref{eq:12}) can be represented in the form
\begin{equation}
t=\frac{t_{0}}{\sqrt{\sigma +1}}\left[ x(t) \mathcal{F}\left( \frac{x^{3}(t)}{x^{3}_{s}}
\right) -\mathcal{F}\left(\frac{1}{x^{3}_{s}}\right) \right],
\label{eq:13}
\end{equation}
where $\mathcal{F}(z) =F\left( \frac{1}{3},\frac{1}{2};\frac{4}{3};z \right)$ and the latter is the hypergeometric
function. Substituting in equation~(\ref{eq:13}) $x(t) =x_{s}$ we obtain the corresponding stopping time as a function
of the magnetic field and the plasma kinetic energy
\begin{eqnarray}
t_{s}=\frac{t_{0}}{\sqrt{\sigma +1}}\left[C \left( \frac{\sigma +1}{\sigma }\right) ^{1/3}
-\mathcal{F}\left(\frac{\sigma }{\sigma +1}\right) \right] .
\label{eq:14}
\end{eqnarray}
Here $C= \mathcal{F}(1)=\sqrt{\pi }\Gamma \left( \frac{4}{3}\right)/\Gamma \left( \frac{5}{6}\right)\simeq 1.4$
is a constant. At vanishing
($\sigma \ll 1$) and very strong ($\sigma \gg 1$) magnetic fields the stopping
time becomes $t_{s} \simeq CR_{m}/v_{m} =C(a_{0}/v_{m}) \sigma^{-1/3}$, $t_{s} \simeq 2R_{m}^{3}/3v_{m }a_{0}^{2}%
=(2a_{0}/3v_{m}) \sigma^{-1}$, respectively, where the radius $R_{m}=(6W_{0}/H_{0}^{2} )^{1/3}$ is obtained by
equating the initial kinetic energy $W_{0}$ of an initially spherical plasma cloud to the energy of the magnetic
field that it pushes out in expanding to the radius $R_{m}$. It is worth mentioning that in the case of weak
magnetic field, $\sigma \ll 1$, and at vanishing thermal pressure ($\beta =0$) the stopping time does not depend
on the initial plasma radius, $t_{s}\sim (M/v_{m} P_{\mathrm{mag}})^{1/3}$.

We now turn to the general equations determined by (\ref{eq:7}) and (\ref{eq:11}). At the initial stage of
plasma expansion ($t\ll t_{0}$) from these equations we obtain
\begin{equation}
x(t) \simeq 1+ \frac{v_{m}t}{a_{0}} +\frac{3}{4}h \left(\frac{t}{t_{0}}\right) ^{2} ,
\label{eq:15}
\end{equation}
where $h =\beta (\gamma -1)-\kappa$, $\kappa =\frac{\alpha}{3}\eta Q'(\eta)$ and $\kappa =\sigma$ for the dipole and
homogeneous magnetic fields, respectively. Here the prime indicates the derivative with respect to the argument. Thus
at the initial stage the plasma cloud may get accelerated or decelerated
depending on the sign of the quantity $h$ (in other words on the relation between thermal and magnetic pressures).
For instance, in the homogeneous magnetic field the acceleration occurs when $p_{\max}>p_{c}$, where
\begin{equation}
p_{c}= \frac{4}{3\sqrt{\pi}} \frac{\Gamma\left(\frac{5}{2}+s\right)}{\Gamma\left(1+s\right)} P_{\mathrm{mag}}
\label{eq:xx}
\end{equation}
(i.e. at $h>0$) and continues until $x(t)$ reaches some value $x_{c}>1$ given by $x_{c} =(p_{\max}/p_{c})^{1/3\gamma}$.
The time interval $0\leqslant t<t_{c}$ of the acceleration is determined from the equation of motion
(\ref{eq:11}). The critical radius $x_{c}$ and time $t_{c}$ correspond to the beginning of plasma deceleration.
Further plasma motion at $t >t_{c}$ is an expansion with slowing-down velocity. It ends up at the turning point
which corresponds to the maximum of expansion, $U(x_{s}) =1$. However, in the opposite case of the low thermal
pressure with $p_{\max}< p_{c}$ the plasma systematically get decelerated in the whole time interval of its dynamics.

\begin{figure}[tbp]
\includegraphics[width=130.0mm]{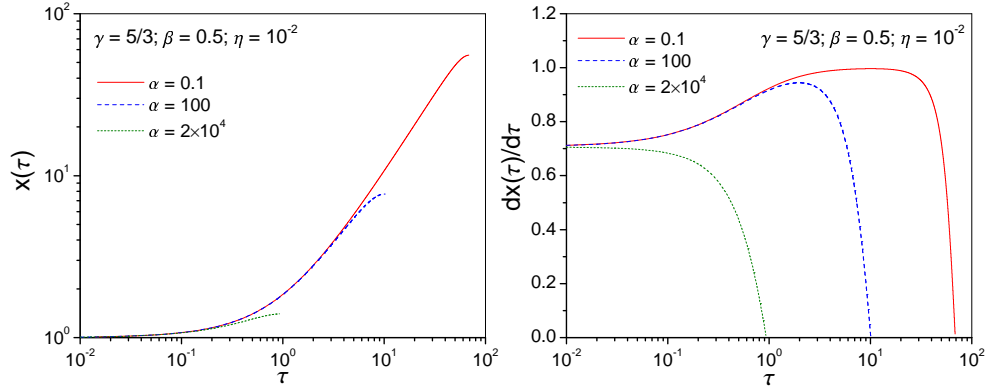}
\caption{The dynamics of the plasma cloud expanding in a dipole magnetic field. Shown are
the scaled radius $a(t)/a_{0}$ (left panel) and the velocity $\dot{a}(t)/u_{m}$ (right panel) of the plasma
boundary vs time (in units of $t_{0}$) at $\gamma =5/3$, $\beta =0.5$, $\eta =10^{-2}$, $\theta_{p} =0$ and
for $\alpha =0.1$ (solid lines), $\alpha =100$ (dashed lines) and $\alpha =2\times 10^{4}$ (dotted lines).}
\label{fig:1}
\end{figure}

A characteristic stopping time of plasma motion up to the full stop at the turning point is given by the integral
of the equations~(\ref{eq:7}) and (\ref{eq:11})
\begin{equation}
\tau_{s}= \int_{1}^{x_{s}} \frac{dy}{\sqrt{1-U(y)}} \simeq 2\sqrt{\frac{x_{s}-1}{U'(x_{s})}} .
\label{eq:16}
\end{equation}
Calculating time $t_{s}$ needed for plasma to reach this point, one can simplify the integrand taking into account
that the main contribution comes from the vicinity of upper limit of integration. This approximation is expressed
by the second part of equation~(\ref{eq:16}). In the case of weak and homogeneous magnetic field this yields universal
expressions, $t_{s} \sim (M/u_{m} P_{\mathrm{mag}})^{1/3}$ and $a_{s} \sim u_{m} t_{s}$. It is worth mentioning that
in the case of weak magnetic field the stopping time and length do not depend on the initial plasma radius but depend
on the thermal pressure (or temperature) (cf. these relations with those obtained above). At very strong magnetic fields,
$a_{s}\simeq a_{0}+(1/2)v_{m} t_{s}$ and $t_{s} \sim Mv_{m}/a^{2}_{0} P_{\mathrm{mag}}$, and the stopping characteristics
of the plasma essentially depend on the initial radius but are now independent on the thermal pressure. The similar
estimates can be found for the dipole magnetic field. However, we note that the latter case significantly differs
from the homogeneous field situation considered above. Since in the vicinity of the dipole the magnetic field is arbitrary
large the stopping length cannot naturally exceed $r_{0}$ for any thermal energy of the plasma ($x(t)<1/\eta$ in
(\ref{eq:7})). For a weak magnetic field this simply yields $a_{s} \simeq r_{0}$ and $t_{s} \simeq r_{0}/u_{m}$.

\begin{figure}[tbp]
\includegraphics[width=130.0mm]{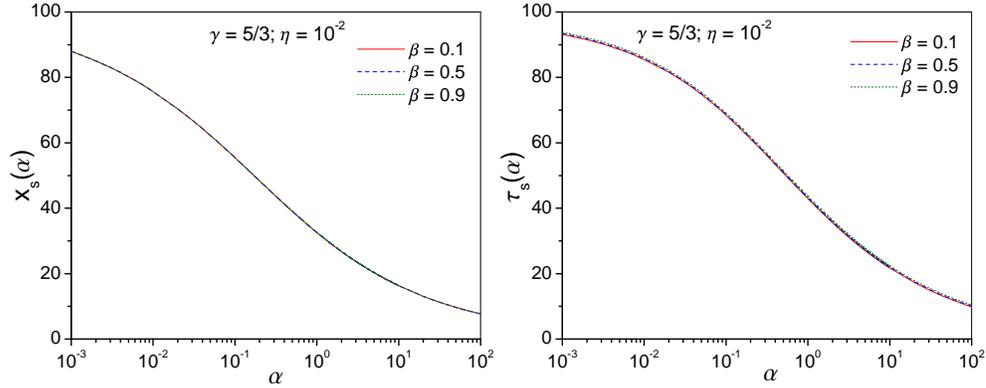}
\caption{The normalized stopping length $a_{s}(\alpha)/a_{0}$ (left panel) and the stopping time
$t_{s}(\alpha)/t_{0}$ (right panel) of the plasma cloud expanding in a dipole magnetic field vs the normalized
dipole magnetic field $\alpha $ at $\gamma =5/3$, $\eta =10^{-2}$, $\theta_{p} =0$ and for $\beta =0.1$ (solid
lines), $\beta =0.5$ (dashed lines) and $\beta =0.9$ (dotted lines).}
\label{fig:2}
\end{figure}

As an example in figure~\ref{fig:1} we show the results of model calculations for the normalized radius $a(t)/a_{0}$
(left panel) and the velocity $\dot{a}(t)/u_{m}$ (right panel) as a function of time (in units of $t_{0}$) at
$\gamma =5/3$, $\beta =0.5$, $\eta =10^{-2}$, $\theta_{p} =0$ and for different values of the parameter $\alpha$.
In this figure the dimensionless strengths $\alpha$ of the dipole magnetic field
are chosen such that the coefficient $h$ in equation~(\ref{eq:15}) is positive, $h>0$, for solid and dashed lines and
negative, $h<0$, for dotted lines. From the right panel of figure~\ref{fig:1} it is seen that at $h>0$ (solid and dashed
lines) there is a short initial period of acceleration, $0\leqslant t\lesssim t_{0}$, when the plasma boundary is
accelerated according to equation~(\ref{eq:15}). During this period (which is only weakly sensitive to the magnetic field
strength) the dimensionless radius $a(t)/a_{0}$ increases up to $2-3$, and at $t_{0}\lesssim t<t_{c}$ almost all initial
total energy $W_{0}$ is transferred into kinetic energy of free radial expansion at constant velocity $\sim u_{m}$. As
expected (see above) the time $t_{c}$ is reduced with increasing the strength of the magnetic field and the free
expansion period is shorter for larger $\alpha$. The further increasing the strength of the magnetic field (figure~\ref{fig:1},
dotted line) results in a plasma dynamics with systematically slowing-down velocity.

For the same set of the parameters $\gamma$, $\eta$ and $\theta_{p}$ in figure~\ref{fig:2} it is shown the normalized
stopping length (left panel) and the stopping time (right panel) of the plasma cloud as a function of the dimensionless
strength $\alpha$ of the dipole magnetic field for some values of the normalized plasma thermal pressure $\beta$. It
is seen that the stopping length and time decrease with the strength of the magnetic field and practically are not
sensitive to the variation of the plasma thermal pressure.

Note that at otherwise unchanged parameters the strength of the dipole magnetic field is maximal at the orientation
$\theta_{p} =0$ and monotonically decreases with $\theta_{p}$. For instance, the strength $H_{0}(0)$ of the dipole
magnetic field at the center of the plasma cloud is reduced by a factor of $2$ by varying the dipole orientation from
$\theta_{p} =0$ to $\theta_{p} =\pi /2$. Therefore the effect of the magnetic field shown in figures~\ref{fig:1} and
\ref{fig:2} is weakened at the orientation $\theta_{p} =\pi /2$ of the dipole. In particular, this results in a larger
stopping lengths and times than those shown in figure~\ref{fig:2}.

In this paper we have assumed that the self-similar solutions (\ref{eq:2})-(\ref{eq:4}) and hence the shape of the
plasma remain isotropic during plasma expansion. Such assumption is not evident since the distribution of the magnetic
pressure on the plasma surface is anisotropic in general. Thus one can expect strong deformation of the initially
spherical plasma surface. Consider briefly this effect assuming constant magnetic field $\mathbf{H}_{0}$. Taking
into account the effect of the induced magnetic field the total magnetic pressure on the plasma surface is
$P_{m}(\theta ) =(9/4)P_{\mathrm{mag}} \sin^{2}\theta$, where $\theta$ is the angle between the radius vector
$\mathbf{r}$ and $\mathbf{H}_{0}$ \cite{rai63,ner06}. The magnetic pressure vanishes at $\theta =0,\pi$ and is maximal
at $\theta =\pi /2$. Therefore, one can expect that the spherical shape of the
plasma is deformed into the ellipsoidal one which is elongated along the magnetic field lines. This effect
is supported by the numerical simulations \cite{osi03,ner09}. Thus, the isotropic self-similar model considered
here is valid as long as the deformation of the plasma shape is much smaller than the typical length scale for the
plasma flow. However, while the solution (\ref{eq:2})-(\ref{eq:4}) is violated in the case of strong anisotropy
the plasma dynamics can be described (at least qualitatively) by equations (\ref{eq:11})-(\ref{eq:16}) replacing
the isotropic magnetic pressure $P_{\mathrm{mag}}$ by the angular dependent one $P_{m}(\theta )$. The expansion
radius $a(t,\theta )$ as well as the stopping length $a_{s}(\theta )$ and time $t_{s}(\theta )$ should be also
anisotropic in this case.

\section{Conclusions}
\label{sec:3}

An analytical self-similar solution of the radial expansion of a spherical plasma cloud in the presence
of a dipole or homogeneous magnetic field has been obtained. The analysis of the plasma expansion into
ambient magnetic field shows that there are processes of acceleration, retardation and stopping at the
point of maximum expansion that are very distinct and separated in space and time.
The scaling laws obtained are, in general, the functions of two dimensionless parameters, $\alpha$ (or
$\sigma$ for constant magnetic field) and $\beta$, which can be varied by means of the choice of the
external magnetic field, the thermal pressure and the initial energy of the plasma. It allows to test
the different regimes of the plasma dynamics in a wide range of external conditions.

We expect our theoretical findings to be useful in experimental investigations as well as in numerical
simulations of the plasma expansion into an ambient magnetic field (either uniform or nonuniform).
One of the improvements of our present model will be the derivation of the dynamical equation for the plasma
surface deformation. In this case it is evident that the problem is not isotropic with respect to the center
of the plasma cloud ($\mathbf{r} =0$) and a full three-dimensional analysis is required. A study of this and
other aspects will be reported elsewhere.

\ack This work was made possible in part by a research grant (Project PS-2183) from the Armenian National
Science and Education Fund (ANSEF) based in New York, USA. Also this work has been supported by the
State Committee of Science of Armenia (Project No.~11-1c317).

\end{document}